\documentclass[twocolumn,showpacs,preprintnumbers,amsmath,amssymb]{revtex4}

\usepackage{graphicx}
\usepackage{dcolumn}% Align table columns on decimal point
\usepackage{bm}% bold math

\newtheorem{theorem}{Theorem}
\newtheorem{acknowledgement}[theorem]{Acknowledgement}

\begin{document}
%\draft

\title{Ferromagnetic materials in the zinc-blende structure}

\author{B. Sanyal, L. Bergqvist and O. Eriksson} 
\affiliation{Department of Physics, Uppsala University, Uppsala, Sweden}

\date{\today}

\begin{abstract}
New materials are currently sought for use in spintronics applications. Ferromagnetic materials
with half metallic properties are valuable in this respect. Here we present the electronic
structure and magnetic properties of binary compounds consisting of 3d transition metals and
group V elements viz. P, Sb and As in the zinc-blende structure. We demonstrate that compounds of V, Cr and Mn show
half metallic behavior for appropriate lattice constants. 
 By comparing the total energies in the ferromagnetic
and antiferromagnetic structures, we have ascertained that the ferromagnetic phase is stable over 
the antiferromagnetic one. Of the different compounds studied, the Cr based systems exhibit the strongest
interatomic exchange interactions, and are hence predicted to have the highest critical temperatures.  
Also, we predict that VAs under certain growth conditions should be a semiconducting ferromagnet.
Moreover, critical temperatures of selected half metallic compounds have been estimated from mean field theory and
Monte Carlo simulations using parameters obtained from a {\it ab-initio} non-collinear, tight binding linearized
muffin-tin orbital method.
>From a simple model, we calculate the reflectance from an ideal MnAs/InAs interface considering
the band structures of MnAs and InAs.
Finally we present results on the relative stabilities of MnAs and CrSb compounds in the NiAs and zinc-blende
structures, and suggest a parameter space in substrate lattice spacings for when the zinc-blende structure 
is expected to be stable. 

\end{abstract}

\vspace{20mm}

\maketitle

%\narrowtext
\section{Introduction}
For magnetoelectronic devices, highly spin polarized ferromagnets 
have a large potential in applications \cite{prinz}. Half metallic
ferromagnets (HMF) are the most desirable materials in this respect. 
These materials have one spin channel that is metallic, 
while the other spin channel has a gap at the Fermi level. 
Hence, 100 $\%$ spin polarization of the conducting electrons is expected from these materials.
Based on band-structure calculations, de Groot {\it et al.} \cite{degroot} 
first predicted half metallic behavior of C1$_b$ type Heusler alloys of NiMnSb and PtMnSb. 
Since then, there have been predictions of many half metallic materials. 
A practical aspect of these materials is the recent breakthrough 
utilizing the spin of electrons in existing semiconductor electronic devices.
To manipulate spins of electrons inside the semiconductors, one first needs to inject spins
inside the semiconductors. Spin injection can be done either from ferromagnetic metals or
from diluted magnetic semiconductors (DMS) doped with magnetic ions. 
 Ohno {\it et al.} \cite{holeinject} successfully injected spin polarized holes from a p-type 
 III-V DMS to a semiconductor. From theoretical calculations of magnetotransport, Egues \cite{egues} 
proposed a spin filter  
for spin-dependent optoelectronics applications. Later on, Fiederling {\it et al.} \cite{fiederling} 
successfully injected spin polarized electrons from an n-type II-VI semiconductor.
 However, according to Schmidt {\it et al.} \cite{schmidt}, the spin injection from a metal to a semiconductor 
is practically impossible due to the huge conductivity mismatch between a metal and a
semiconductor. The use of a DMS is instead rather 
advantageous. Mn doped GaAs is a prototype of this class of systems but the highest available
Curie temperature is only 110 K \cite{ohno} due to the limited solubility of Mn in GaAs.
The use of other half metallic systems e.g. CrO$_{2}$ and La$_{0.7}$Sr$_{0.3}$MnO$_{3}$ 
is also restricted due to lower Curie temperature. Very recently, there have been reports on room 
temperature ferromagnetism in Mn doped GaP compounds \cite{mngap}. Hence, the quest for new half metallic 
materials with room
temperature ferromagnetism continues.

 Recently, there have been efforts to grow binary compounds in the zinc-blende structure using
molecular beam epitaxy. One of the advantages of these materials is that the zinc-blende structure is
 maintained when grown on a zinc-blende semiconductor substrate. Akinaga {\it et al.} \cite{akinaga} 
grew CrAs on GaAs substrates and found CrAs to be a ferromagnet
having a Curie temperature higher than 400 K. The experimental results were complemented by {\it ab-initio}
electronic structure calculations showing that the magnetic moment was 3 $\mu_{B}$. Also CrSb was
grown in the zinc-blende structure \cite{crsb}. These binary compounds do not exist naturally in the
zinc-blende structure but the non equilibrium growth by MBE has facilitated the creation of these
compounds with novel magnetic properties.

Sanvito and Hill \cite{sanvito} examined the ground state properties of MnAs in a NiAs and a zinc-blende
structures by first principles pseudopotential calculations.
 Shirai \cite{shirai} performed electronic structure calculations for the zinc-blende compounds of
 transition metal (TM) and As. The early TM-As compounds were shown to be ferromagnets and the later compounds were
antiferromagnets. Also, electronic structure calculations have been performed for different
 crystal structures of MnP, MnAs and MnSb by Continenza {\it et al.} \cite{conti}.
Galanakis \cite{galan} examined the half metallicity of the bulk and surface of zinc-blende
CrAs in various lattice constants by electronic structure calculations.  
 A full potential screened Korringa-Kohn-Rostoker method was used. 
 Galanakis concluded that a Cr terminated
  CrAs (001) surface remains
 half metallic both for the experimental 
 lattice constants of GaAs and InAs. On the contrary, an As terminated
 surface loses its half metallicity due to the presence of the surface dangling bond states.

 The motivation of this work is to design new materials in a crystal structure that is suitable for 
spin electronics applications and provide a guideline for experimental work that attempt to grow
these materials.
The paper is organized as follows : In section 'Computational details', we describe in detail 
the computational part. In the
results section, a subsection 'Electronic structure and magnetism' 
describes the electronic structure and magnetism of the binary
compounds considered followed by the discussions about the 
 interatomic exchange interactions and critical temperatures 
in some of the systems
studied. 
Finally, structural stabilities of MnAs and CrSb
compounds in various structural phases are presented.  

\section{Computational details}
A zinc-blende structure has two fcc sublattices displaced from each other along the body diagonal by
 a vector ($\frac{1}{4} \frac{1}{4} \frac{1}{4}$). For calculations in the ferromagnetic phase, 
the unit cell used in the calculations is the same as in
the fcc lattice and has 1 TM at the (000) and 
the group V atom at the ($\frac{1}{4} \frac{1}{4} \frac{1}{4}$)
position. 
Calculations have been performed by an {\it ab-initio} plane wave pseudopotential
code (VASP) \cite{vasp}. Pseudopotentials generated using the  
Projector augmented wave (PAW) \cite{paw} method were used throughout the calculations.
These pseudopotentials are suitable for transition metals (TM) and the calculations yield accurate results
\cite{tmgaas}.
 A kinetic energy cut-off of 600 eV was
used for the plane waves included in the basis set. 
Perdew-Wang GGA (generalized gradient approximation) exchange -correlation \cite{pw} was
used. According to the calculations of Continenza {\it et al.} \cite{conti}, 
GGA is essential for obtaining accurate equilibrium volume and magnetic moments. 
A 20x20x20 k-points grid was used in the Monkhorst Pack scheme \cite{mp} 
which yielded 770 k points in the irreducible Brillouin zone and 
subsequently used for the Brillouin zone integration within the
tetrahedron method. Local properties such as local density 
of states and local magnetic moments
were calculated by projecting the wave functions into spherical 
harmonics as described in ref. \cite{eichler}.
 The values of the experimental lattice constants 
 used in these calculations were taken from 
  Ashcroft and Mermin \cite{ash}. 
Calculations were done 
for the TM-V compounds for the 
lattice constants of the III-V semiconductors having the same Group V element.  
In summary, we considered lattice constants of GaAs, GaP, GaSb, InAs, InP and InSb compounds
which covers a wide range (5.45 \AA \ for GaP to 6.48 \AA \ for InSb).

The pseudopotential calculations have been 
complemented \cite{footnote} with calculations based on non-collinear 
tight-binding linear muffin-tin orbital method (TB-LMTO-ASA)\cite{andersen} 
including combined correction terms in the one-electron Hamiltonian\cite{skriver}.  
The generalized gradient approximation (GGA) were used for the exchange-correlation 
potential\cite{pbe}. Equal Wigner-Seitz sphere radii were used for all atoms, 
as well as for the empty spheres representing the two types of interstitial sites 
in the zinc-blende structure. A dense 27x27x27
k-point grid was used in the Brillouin zone integration 
and to speed up convergence each energy eigenvalue was 
associated with a Gaussian of width 85 meV.
Spin wave excitations have been studied using 
 the frozen-magnon approach\cite{rosengaard,halilov,pajda}. 
Then the total energy $E(\bf{q},\theta)$ is calculated for 
spiral magnetic structures where each magnetic moment is defined by the Euler angles
 
\begin{equation}
\theta_i = const   ;    \phi_i=\textbf{q} \cdot \textbf{R}_i,
\end{equation}
where $\theta_i$ and $\phi_i$ are the polar and azimuthal angles, $\textbf{R}
_i$ the position of ion $i$ and $\textbf{q}$ is the spiral propagation 
vector. Spin spirals possess a 
generalized translational symmetry\cite{sandratskii}, 
which allow for calculations in the chemical 
unit cell without need for a supercell. It can be shown, that the true
spin wave excitation energies are related to the spiral energies through the relation
 
\begin{equation} \label{spinwave}
\omega(\textbf{q})=  \frac{4}{M} \frac{\Delta E(\textbf{q},\theta)}{sin^2\theta},
\end{equation}
where $\Delta E(\textbf{q},\theta) = E(\textbf{q},\theta)-E(0,\theta)$. 
The last equation is valid for one atom per cell.
Thus the formula is correct if one consider
the magnetic response from the transition metal only and neglect the small induced magnetic moment on the group V element.
In our calculations, we use an angle $\theta = 20^{\circ}$ 
and the spiral energies were obtained by employing 
the magnetic force theorem. 
We tested
other angles as well as self-consistent
calculations and found that the differences in total energy is within a few percent.

In order to obtain 
thermodynamic properties, the spin wave 
excitation energies were mapped to an effective Heisenberg Hamiltonian of classical spins
 
\begin{equation} \label{heisenberg}
H =- \sum_{i \ne j}     J_{ij} \textbf{e}_i \cdot \textbf{e}_j ,
\end{equation}
where $J_{ij}$ are exchange interactions between the transition metal atoms. 
The critical temperatures $T_c$ were estimated both from 
mean field theory\cite{ash} (MFA) and Monte Carlo simulations (MC). 
 For the MC simulations, we used the standard single flip Metropolis algorithm 
and the critical temperature were obtained from the 
'cumulant crossing method' \cite{Binder}.
In this method, the reduced fourth order cumulant of the order parameter (magnetization) $U_L$, defined as
\begin{equation}
U_L = 1- \frac{\langle M ^4 \rangle}{3 \langle M ^2 \rangle ^2},
\end{equation}
is calculated for different sizes of the simulation box.
The curves of $U_L$ have a common intersection point at a
fixed point $U^*$, which corresponds to the critical point.
Hence, a value of $T_c$ is obtained from the intersection point of $U_L$ for different lattice sizes.

\section{Results and discussions}
\subsection{Electronic structure and magnetism}
The zinc-blende lattice introduces a tetrahedral crystal field under which the transition metal
 atomic $d$-states are split into triply degenerate $t_{2}$ and doubly degenerate $e$ states.
 $e$ states lie lower in energy than $t_{2}$ states in a tetrahedral crystal field in contrast to
 the octahedral crystal field.
  The $t_{2}$ states hybridize with the anion $p$-states. As a result, bonding and
 antibonding states appear. Also, depending on the exchange splitting, the bonding and antibonding states
 may lie in different positions for different spin channels. $e$ states are of non-bonding character
and do not take part in the hybridization.

%% FIG 1
\begin{figure}
\resizebox{9.0cm}{!}{\includegraphics[angle=270]{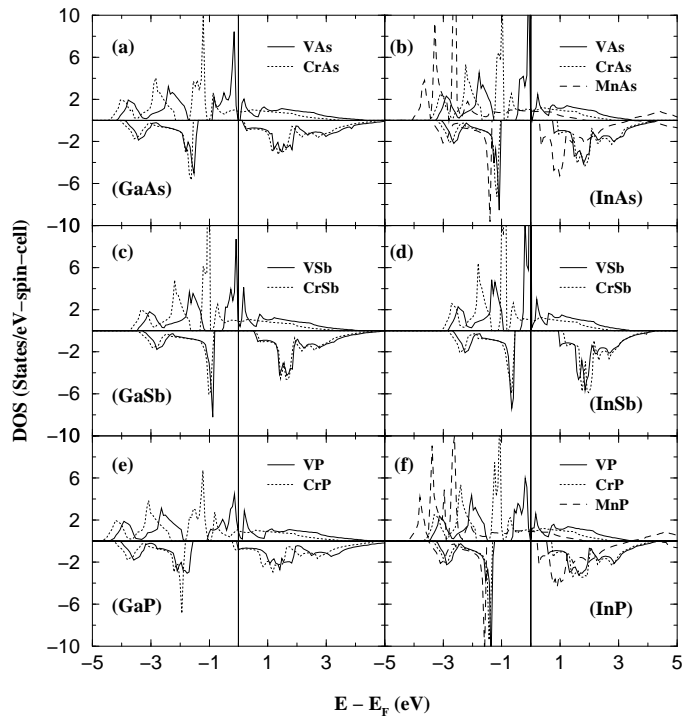}}
\caption{Spin resolved total (sum over atoms and orbitals) density of states (DOS) 
 calculated for the unit cell. In (a), we 
show the DOSs for VAs and CrAs in the GaAs lattice constant, in (b),
VAs, CrAs and MnAs in the InAs lattice constant, in (c), VSb and CrSb in the 
GaSb lattice constant, in (d), VSb and
CrSb in the InSb lattice constant, in (e), VP and CrP in the GaP 
lattice constant and in (f), VP, CrP and MnP in the InP
lattice constant. 
The energies are shifted with respect to the corresponding Fermi levels and
energies are in electron volts.}
\end{figure}

In Fig. 1(a-f), we present the spin resolved DOSs of V, Cr and Mn compounds having the lattice constants
of GaAs, InAs, GaSb, InSb, GaP and InP. In figs. (a), (b), (c), (d) and (f), all the compounds show
half metallic behavior since there is no state at the Fermi level for the spin down DOS. 
The only exception is the case depicted in fig. 1(e) which doesn't exhibit 
half metallic ferromagnetism as the Fermi level cuts both the spin bands. For the lower lattice constant of 
GaP, the $p-d$ overlap and hybridization of the spin down states broadens the band to cut the Fermi level.  
 Also, as the width
of the gap in the spin down DOS is quite large, the Fermi level stays within the gap with the variation in
the lattice constant. It is worthy to be mentioned here that these compounds can be regarded to be 
the derivatives of diluted magnetic semiconductors, e.g. Mn$_{x}$Ga$_{1-x}$As for a Mn doped GaAs system
having $x$=1. Several first principles calculations \cite{sanprb01,schultess,zhao,pashaprl}
confirmed that the total spin magnetic moment of Mn doped GaAs for various 
concentrations of Mn is 4 $\mu_{B}$/cell with a half metallic nature.
 It was also shown \cite{tmgaas} that Cr doped GaAs is a half metallic 
ferromagnet with a total spin magnetic moment of 3 $\mu_{B}$/cell. 
 We find here a similar trend for
the extreme limit at a high TM concentration. 
A systematic variation in the DOSs of V, Cr and Mn is seen. As the number of 
valence electrons increases by going from V to Mn, the main features of the DOSs 
shift to lower energies relative to the Fermi level. This can be seen from
figs. 1 (b) and 1 (f). The exchange splitting is maximum for Mn and here one finds a maximum separation
between the spin up and spin down bands.

A close inspection of
Fig. 1(a) reveals that VAs behaves like a 
magnetic semiconductor for the lattice constant of GaAs, i.e. the Fermi
level passes through a gap for both spin up and spin down channels. We varied the lattice constant of VAs
around the GaAs lattice constant and found that the small gap is sensitive to the lattice spacing. 
Only in a very small region
around the GaAs lattice constant, do we find that the gap persists. This is an interesting result since 
it is one of quite few ferromagnetic
semiconductors found in a stoichiometric composition.
In addition, it is to be noted that unlike CrAs, MnAs doesn't show half metallic behavior 
for the GaAs lattice constant. 
It is instead half metallic for the InAs lattice constant (see inset of Fig. 6(b)). 

In Table I, we present the 
total magnetic moment/cell for all the compounds considered. The half metallic
ferromagnetic moments are marked in bold. Half metallicity may be monitored from the integer moments
of the unit cell.
It is seen that from all the combinations, VAs and CrAs are 
HMFs for both GaAs and InAs lattice constants. VSb and CrSb are also half metallic for GaSb and InSb lattice 
constants. VP, CrP and MnP are only HMFs at the InP lattice constant. This shows that the lattice constants
 and the resulting electronic structures 
 play an important role in determining half metallicity, a finding that is also evident from Fig. 1. 
The moments listed in Table I are consistent with the calculated moments of Cr and Mn impurities in a
GaAs host, if one neglects the effect of other defects e.g. As antisites \cite{pashaprl}.
To be more specific we note that Mn as an impurity in GaAs carries a
magnetic moment of 4 $\mu_B$/Mn atom in all theoretical first principles
 calculations based on local density approximation (LDA) that
neglect other lattice defects such as As antisites. If one considers As
antisite defects, or Mn interstitial atoms that couple
antiferromagnetically to the Ga substituted Mn atoms, the effective
moment becomes $\sim$2.4 $\mu_B$/Mn atom in agreement with experimental
data \cite{pashaprl}. For the currently investigated materials only a
limited set of experimental data have been published. However, CrAs has
been synthesized in the zinc-blende structure and the magnetic moment
was measured to be 3 $\mu_B$/Cr atom\cite{akinaga} in agreement with the present theory and
previous calculations\cite{akinaga}. Since Mn has one extra electron in
the  spin up band, it is natural that the moment of MnAs is 4$\mu_B$/Mn
atom, which is in agreement with the moment of dilute Mn in GaAs (if one
neglects defects).

The explanation of the half metallicity and the integer moments is similar to that of a III-V 
diluted magnetic semiconductor. In a III-V DMS, a cation
is removed by creating a vacancy containing 3 holes in the valence band.   
 When a transition metal occupies this cation
site, it donates 3 electrons to the anion dangling bonds to saturate the bonding. The remaining unpaired 
d electrons of the TMs are 
responsible for the magnetic moments. Due to the large exchange splitting according to Hund's
rule, the spin up and spin down d bands are separated from each other having different degree 
of hybridization
with the valence bands. Hence, V, Cr and Mn have magnetic moments of 2, 3 and 4 $\mu_{B}$, respectively
from the unpaired spin up d electrons. However, the widths of the $d$-bands are fairly large indicating a substantial
amount of hybridization with the anion-$p$ orbitals.

\vspace{4mm}
\noindent{\bf TABLE I.} Total magnetic moment/cell (in $\mu_{B}$). Each entry corresponds to the binary alloy
consisting of a transition metal from the leftmost column and the group V element in the subsequent columns. The 
bulk semiconductors whose lattice constants are considered for each calculation are shown in suffices.  
\vspace{2mm}

\begin{center}
\begin{tabular}{|l|l|l|l|l|l|l|}
\hline \hline 
                         & As$_{GaAs}$ & Sb$_{GaSb}$ & P$_{GaP}$ & As$_{InAs}$ & Sb$_{InSb}$ & P$_{InP}$ \\
\hline
Ti                       & 0.00        & 0.00        &  0.01     &  0.29       & 0.00        &  0.04      \\         
\hline
V			 & \bf{2.00}   & \bf{2.00}   & 1.90      & \bf{2.00}   & \bf{2.00}   & \bf{2.00}  \\ 
\hline
Cr			 & \bf{3.00}   & \bf{3.00}   & 2.74      & \bf{3.00}   & \bf{3.00}   & \bf{3.00}  \\
\hline
Mn			 & 3.53        & 3.89        & 3.19      & \bf{4.00}   & 4.02        & \bf{4.00}  \\
\hline
Fe			 & 2.36        & 2.69        & 1.06      &  3.20       & 3.26        & 3.15       \\
\hline
Co			 & 0.02        & 0.23        & 0.00      & 1.15        & 1.39        & 0.94	  \\
\hline
Ni			 & 0.00        &  0.02       &  0.01     &  0.04       & 0.02        &  0.11      \\
\hline \hline
\end{tabular}
\end{center}

We also investigated compounds of Ti, Fe, Co and Ni, and found that they
 do not show half metallic ferromagnetism (see Table I). Fe compounds have fairly big moments in
the ferromagnetic phase except for FeP in the GaP lattice constant. CoSb in the InSb lattice constant has the highest 
moment among the Co based compounds. Ti and Ni based compounds do not have prominent ferromagnetism. As only V, Cr and
Mn based compounds show interesting and useful magnetic behavior, we will now concentrate on those compounds only.   

\vspace{4mm}
\noindent{\bf TABLE II.} Total energy differences in eV/Mn atom ($\Delta E = E_{AFM}-E_{FM}$) where $FM$ and
$AFM$ are the 
ferromagnetic and antiferromagnetic arrangements between Mn atoms respectively. The symbol 
'-' is for the compositions
where calculations were not performed. 
\vspace{2mm}

\begin{center}
\begin{tabular}{|l|l|l|l|l|l|}
\hline \hline 
                         & GaAs   & InP    & InAs     & GaSb     & InSb \\
\hline
VAs                      & 0.17   & -      & 0.19     & -        & -   \\         
\hline
CrAs                     & 0.27   & -      & 0.32     & -        & -    \\ 
\hline
VP			 & -      & 0.2    & -        & -        & -    \\
\hline
CrP		         & -      & 0.31   & -        & -        & -     \\
\hline
MnP		         & -      & 0.14   & -        & -        & -     \\
\hline
VSb		         & -      & -      & -        & 0.18     & 0.12  \\
\hline
CrSb	                 & -      & -      & -        & 0.31     & 0.27   \\
\hline \hline
\end{tabular}
\end{center}

\subsection{Interatomic exchange interactions}
In this section, we compare the total energies of the HMFs found from our calculations, in the ferromagnetic (FM) 
and antiferromagnetic (AFM)
phases. For briefness, calculations were only made for selected V, Cr and Mn compounds (see Table II). 
In our calculations, we considered an 8 atom (TM)$_{4}$X$_{4}$ 
cell in a layer geometry in the (001) direction with 2 TM (TM$_{I}$) atoms in the plane
at z=0.0 and 2 TM atoms at z=a/2 (TM$_{II}$), where a is the lattice parameter. Here TM and X
 indicate transition metal (V, Cr and Mn) and the group V element (As, P and Sb) respectively. 
TM$_{I}$ and 
TM$_{II}$ atoms can couple ferromagnetically or antiferromagnetically. 
%% FIG 2
\begin{figure}
\resizebox{6.0cm}{!}{\includegraphics[angle=0]{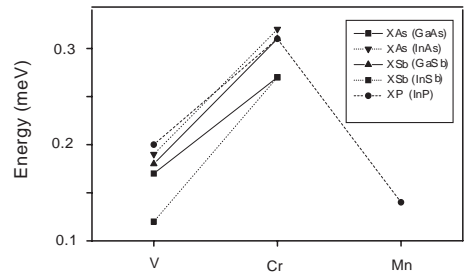}}
\caption{Exchange interactions for V, Cr and Mn compounds.
 Note that the data are given for lattice constants corresponding to
 GaAs, InAs, GaSb, InSb and InP substrates.}
\end{figure}
The energy differences are
given in Table II and are also plotted in Fig. 2. The energy differences, which reflects the interatomic 
exchange coupling strengths are fairly large and are 
found to always stabilize ferromagnetism. 
 From Fig. 2 and Table II, it is clear that the strongest exchange interaction is found for the Cr based compound, 
where the energy difference is quite 
appreciable, indicating a high Curie temperature. An
inspection of the DOSs for the FM and the AFM structures reveal that for the AFM structure, the 
DOS at the Fermi energy is non zero
for both spin up and spin down 
bands whereas in the FM alignment, only spin up states exist at the Fermi level. This
stabilizes the FM alignment, 
since the eigenvalue-sum in general lowers the total energy when electron states are removed
from the Fermi energy. 

%% FIG 3
%%\begin{figure}[!hbp]
\begin{figure}
\includegraphics[width=0.45\textwidth,height=0.2\textheight]{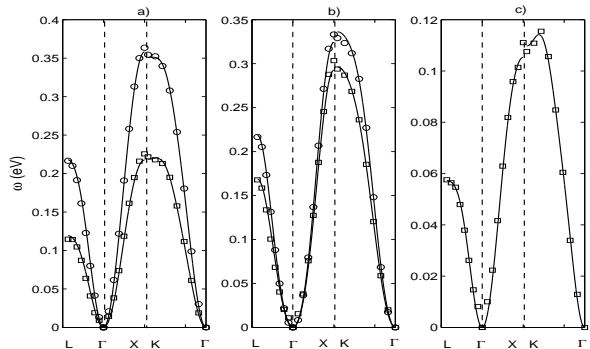}
\caption{Spin wave spectra along high-symmetry lines 
in the Brillouin zone of a) VAs b) CrAs c) MnAs. 
Circles are calculated values for the lattice constant of GaAs and 
squares are for the lattice constant of InAs. Solid lines represent the fitted data.}
\end{figure}
 
In this section, we also present results of the calculations of critical 
temperatures using spin wave spectra and statistical mechanics. 
We have calculated $T_c$ for VAs, CrAs and MnAs for 
lattice constants of GaAs and InAs by extracting exchange interactions
from calculated spin wave spectra, which subsequently were used in a 
classical Heisenberg Hamiltonian. 
%% FIG4
\begin{figure}[!hbp]
\includegraphics[width=0.45\textwidth,height=0.25\textheight]{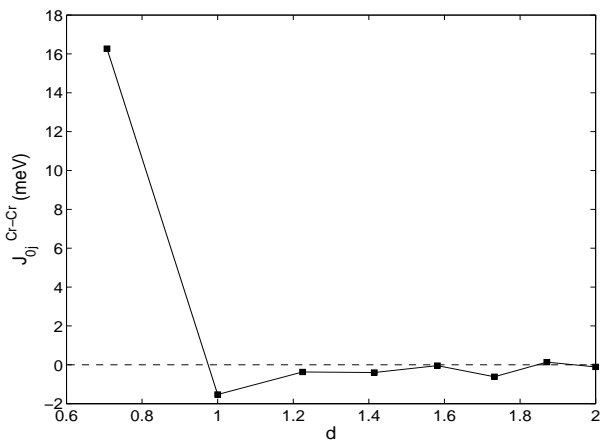}
\caption{Exchange interactions between the Cr atoms in 
CrAs as a function of Cr-Cr distance (in units of GaAs lattice constant).}
\end{figure}
In Fig. 3, the calculated spin wave spectra for the systems are shown. 
MnAs in the lattice constant of GaAs was found to be 
antiferromagnetic and is therefore not shown in the figure. 
The other systems were found to be ferromagnetic and half metallic 
with total magnetic moments of 2.0, 3.0 and 4.0 $\mu_B$, respectively for VAs, CrAs and MnAs. 
In Table.III, calculated values of $T_c$ are shown. 
In the case of CrAs, a very high $T_c$ around 1000 K is 
estimated which could be very promising from an application 
point of view. $T_c^{MC}$
 is smaller than $T_c^{MFA}$ due to the 
fact that fluctuations are absent in the latter. 
Normally, MC simulation gives $T_c$ values closer to experimental values than 
those obtained from mean field theory\cite{rosengaard}. 

\vspace{2mm}
\noindent{\bf TABLE III.} Calculated critical temperature of VAs, 
CrAs and MnAs from the MFA and MC simulations. Values at the left 
correspond to the lattice constant of GaAs and the right correspond to InAs.
\vspace{2mm}

\begin{table}[!hbp]  \label{res1}
\begin{center}
\begin{tabular}{|c| c| c|| c| c|}
\hline
\hline
   & $T_c^{MFA} (K)$ & $T_c^{MC} (K)$ & $T_c^{MFA} (K)$ & $T_c^{MC} (K)$  \\
\hline
VAs & 990 & 830 & 610 & 490 \\
\hline
CrAs & 1320 & 980 & 1100 & 790\\
\hline
MnAs & - & - & 640 & 530 \\
\hline
\hline
\end{tabular} 
\end{center}
\end{table}

As is clear from Table III, the critical temperatures of VAs and CrAs are in general
larger for the substrate with the smallest lattice constant (GaAs). Hence it is clear that
the critical temperatures of VAs and CrAs follow the same trend as that of 
rare-earth systems, where in general a reduced lattice constant results in enhanced
critical temperatures (as for instance measured by an applied external pressure
\cite{rare}). In rare-earths this phenomenon is easily understood from the RKKY interaction,
 which is known to be responsible for the magnetic ordering in these elements. 
By analogy we can say that the presently studied materials (VAs and CrAs) have rare-earth like
characteristics, in the sense that the exchange interaction between localized moments is 
mediated by more diffuse
valence states. 

In fig. 4, we show the exchange interactions ($J_{ij}$) in CrAs as a function of Cr-Cr distance.
It is clear that the first nearest neighbor exchange interaction is the dominant one and is
ferromagnetic in nature. $J_{ij}$ decreases
rapidly with increasing distance between Cr atoms. Further neighbor exchange interactions are
antiferromagnetic in nature and are very weak. Similar kind of observation was reported by
Sandratskii and Bruno \cite{sand} in the case of Mn doped GaAs.

\section{Band structure and calculation of reflectance}

In Fig. 5(a) and (d), we show the band structures of MnAs in the InAs lattice constant for two
symmetry directions viz. $\Gamma$-L and $\Gamma$-X directions respectively. In (b) and (c), we
show the same for bulk InAs calculated for the experimental lattice parameter. 
Our calculation does not produce a gap at the $\Gamma$ point, so the top of the 
valence band and the bottom of the conduction band merge with each other.   
This is a well known limitation of using LDA or GGA within density functional theory.
 For MnAs, the spin up bands cross the Fermi level for the two symmetry directions
 shown the figure and also for other directions (omitted in the figure). There is gap in the
spin down channel. As described above, the $e$ bands are dispersionless as they do not mix with others.
The antibonding $t_{2}$ bands show large dispersion as they are hybridized strongly with the As $p$ bands. 

%% FIG 5
\begin{figure} [!hbp]
\resizebox{8.5cm}{!}{\includegraphics[angle=270]{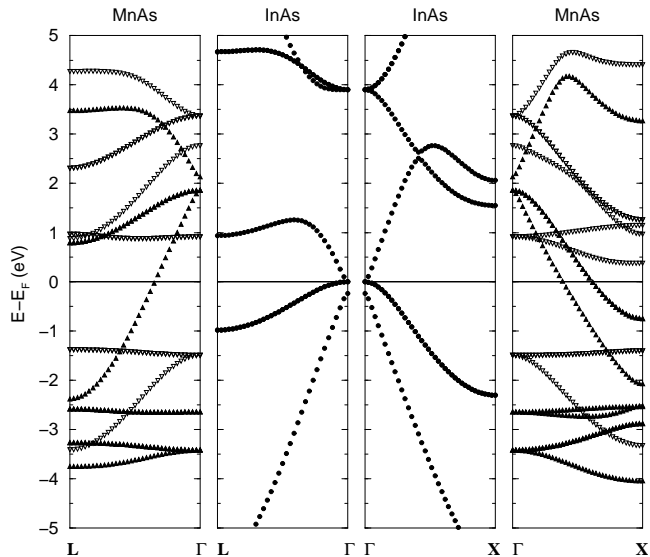}}
\caption{Band structure of MnAs and InAs in two symmetry directions. 
For MnAs, the filled up triangles and empty down triangles indicate spin up and spin down bands respectively. 
The horizontal lines indicate the Fermi level of MnAs and the valence band maximum of InAs.}
\end{figure}

We now give a rough estimate of the reflectance and transmittance from a 
 ferromagnetic metal/semiconductor interface from the band matching of the individual species.
We use a simple free electron model, that was used also by Kilian and Victora \cite{kilian2000}. For an 
electron traveling through a one-dimensional potential step, the reflectance is given by
\begin{equation}
R = \frac{(k_{fm}-k_{sc})^{2}}{(k_{fm}+k_{sc})^{2}} ~,
\end{equation}
where $k_{fm}$ and $k_{sc}$ are the wave vectors in the ferromagnetic metal and semiconductor
respectively.  As the spin up bands are the only states available at the 
Fermi energy, we consider the following for
 the spin up electrons only. 
 For InAs, as the conduction band minimum is at the $\Gamma$ point, the lowest energy levels available
to accept spins from the metal are at the $\Gamma$ point. 
We calculate $R[100]$, $R[110]$ and $R[111]$ for [100]
($\Gamma$-$X$), [110]($\Gamma$-$K$)
and [111]($\Gamma$-$L$)  directions, respectively. 
It is known from experiments that for e.g. an interface of
Au/{\it n}-InAs, the Fermi level is pinned 0.1 eV above the conduction band minimum. 
We use this information in our model assuming that the properties of the metal do not alter
 strongly the band alignments. Also, the spins within 
($k_{B}T \sim 0.025 eV$) of the MnAs Fermi level are the only ones 
available for transport properties. From this information we
can calculate $k_{sc}$ for the three different directions. 
$k_{fm}$ for the three different directions are available from
the band structures of MnAs directly. Then we calculated the 
 reflectance from equation 1. The calculated 
values are $R[100]$=0.73, $R[110]$ = 0.89 and $R[111]$ = 0.71. This means that the transmissions in these
directions will be rather small. However, one should bear in mind that in reality the interface is not
 ideal. There are interface states and different scattering mechanisms which can further decrease or increase
the transmission through the interface. A full transport calculation 
taking into account a proper model of the
interface could give a more realistic estimate, and our model should be viewed as an estimate only.

\section{Structural stability}
In this section, we present the relative structural stabilities of these compounds. 
For briefness, we choose to illustrate our conclusion only for 2 out of the 9 HMFs shown in 
Table I, i.e. CrSb and MnAs. The conclusions reached for these two compounds should be valid for all 
the HMFs listed in Table I.
 We  
compare the total energies of two crystal structures, hexagonal NiAs (that is the low temperature 
structure) and the zinc-blende (ZB) structure. 

%% FIG 6
\begin{figure}[!hbp]
\resizebox{8.0cm}{!}{\includegraphics[angle=270]{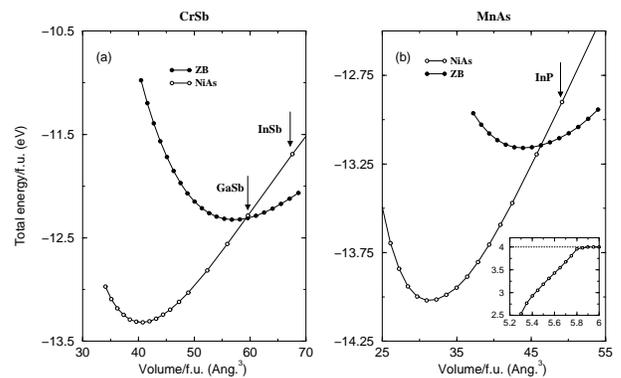}}
\caption{Total energy vs. volume for (a) CrSb in the zinc-blende and the NiAs structure (b) MnAs in the 
zinc-blende and the NiAs structure with (inset) variation of total magnetic moment/cell with lattice 
constant for zinc-blende MnAs.}
\end{figure}
From Fig. 6, it is 
evident that for both MnAs and CrSb, the NiAs structure is more stable than the ZB structure. Similar results 
were obtained for MnAs by Sanvito and Hill \cite{sanvito}.  
This is 
in accordance with naturally occurring compounds of CrSb and MnAs in the 
NiAs structure. However the total energy curves for
 the NiAs structure crosses the total energy curves of ZB structures for certain volumes. This occurs for
a volume of 46.3 $ \AA\ ^{3}$/f.u. in MnAs and 59.6 $ \AA\ ^{3}$/f.u. 
 in CrSb. Incidentally, these volumes are quite close to that of 
GaAs in case of ZB-MnAs and that of GaSb in case of ZB-CrSb
compounds. 
For larger volumes, the ZB structure is more stable than the NiAs structure.
Also, these compounds become half metallic ferromagnets according to the calculations.
Therefore, ZB heterostructures of suitable combinations 
should be possible to use to obtain half metallic materials with a large critical 
temperature. 
 Hopefully this fact can be a useful guideline for experimental work 
in trying to grow heterostructures for spin-electronics applications.
Finally, the variation of magnetic
moment of MnAs with the variation in lattice constant is shown in the 
inset of Fig. 6(b) where it is clear that it becomes half metallic
only near the vicinity of a lattice constant of 6 \AA\ . 

The results presented above are in excellent agreement with the existing ones, wherever 
available. However, there are certain limitations of our approach. One of them is 
the underestimation of band gaps in semiconductors within the use of LDA or GGA. A more appropriate
way is to use GW approximation \cite{gw} 
for these materials which corrects the band gaps extremely well. To our knowledge, no theoretical
calculation exists for diluted magnetic semiconductors using this more complicated theory.
The other issue is the effects of native defects which are always present in the growth processes.
One should also take into account these effects to model a more realistic situation. But we believe that
the results presented in this paper are quite significant as far as trends are concerned and also a 
guideline for future experiments.
  
\section{Conclusion}
In conclusion, we have identified a number of 
possible half metallic ferromagnets in the zinc-blende structure.
Ferromagnetism has been 
shown to be more stable than an antiferromagnetic structure and exchange couplings are 
 calculated to be sufficiently strong
 to produce high Curie temperatures, especially for Cr based compounds.  
 We also give a parameter range in substrate 
lattice constants where MnAs and CrSb are expected to grow and to have half metallic properties. 
Finally, we predict that VAs grown at suitable conditions should be a ferromagnetic semiconductor with a 
large critical temperature.
We hope to motivate 
experimental work in trying to grow such 
materials for future applications. From a theoretical point of view, 
the future scope of this work is to 
investigate the interface properties of these binary compounds and the zinc-blende semiconductors e.g. CrAs
and GaAs. A detailed investigation including structural relaxations of these interfaces is in progress.

\begin{acknowledgement}
This work was done during the stay of one of the authors (B.S.) in Max-lab, Lund, Sweden. 
Support from the Swedish Science Foundation (VR), Swedish Foundation for Strategic
Research (SSF) and the G\"oran Gustafsson foundation are acknowledged.
 Also, we acknowledge support from National Supercomputer Center (NSC).
\end{acknowledgement}


\begin{references}
\bibitem{footnote} We have done general tests to compare ASA and full potential (FP) results. They differ only
by a few percent e.g. in the equilibrium lattice constant and magnetic moments.
 Values of exchange parameters are also very similar. Nearest neighbour exchange interaction
parameters extracted from ASA calculations tally extremely well with those calculated from
total energy differences in FP calculations.
\end{references}
\end{document}